\documentclass[fleqn,usenatbib]{mnras}

\usepackage[T1]{fontenc}

\DeclareRobustCommand{\VAN}[3]{#2}
\let\VANthebibliography\thebibliography
\def\thebibliography{\DeclareRobustCommand{\VAN}[3]{##3}\VANthebibliography}

\usepackage{graphicx}	% Including figure files
\usepackage{amsmath}	% Advanced maths commands
\usepackage{amssymb}	% Extra maths symbols
\usepackage{multirow}
\usepackage[version=3]{mhchem}
 \usepackage{xcolor}
\newcommand{\degK}{\,\mbox{K}}

\newcommand{\kms}{\mbox{\,km s}^{-1}}

\newcommand{\Myr}{\mbox{\,Myr}}
\newcommand{\pc}{\mbox{\,pc}}
\newcommand{\pcc}{\mbox{\,cm}^{-3}}

\title[Molecular filament widths as tracers of accretion]{Different molecular filament widths as tracers of accretion onto filaments}

\author[Gómez, Walsh, \& Palau]{
Gilberto C. Gómez,$^{1}$\thanks{E-mail: g.gomez@irya.unam.mx}
Catherine Walsh,$^{2}$
and Aina Palau$^{1}$
\\
$^{1}$Instituto de Radioastronomía y Astrofísica, Universidad Nacional Autónoma de México, Antigua Carretera a Pátzcuaro \# 8701, \\ Ex-Hda. San José de la Huerta, Morelia, Michoacán, México C.P. 58089\\
$^{2}$School of Physics and Astronomy, University of Leeds, Leeds, UK LS2 9JT
}

\date{Accepted XXX. Received YYY; in original form ZZZ}

\pubyear{2021}

\begin{document}
\label{firstpage}
\pagerange{\pageref{firstpage}--\pageref{lastpage}}
\maketitle

\begin{abstract}
We explore how dense filament widths, when measured using different molecular species,
may change as a consequence of gas accretion toward the filament.
As a gas parcel falls into the filament, it will experience different density, temperature, and extinction values.
The rate at which this environment changes will affect differently the abundance of different molecules. So,
a molecule that forms quickly
will better reflect the local physical conditions a gas parcel experiences than 
a slower-forming molecule.
Since these differences depend on how the respective timescales compare, 
the different molecular distributions should reflect how rapidly the environment changes, i.e.,
the accretion rate toward the filament.
We find that the filament widths measured from time-dependent abundances for \ce{C2H}, CO, CN, CS, and \ce{C3H2}, 
are the most sensitive to this effect. This is because these molecules are the ones presenting also the wider filament widths.
On the contrary, molecules such as \ce{N2H+}, \ce{NH3}, \ce{H2CO}, HNC and \ce{CH3OH} are not so sensitive to accretion and present the narrowest filament widths. We propose that ratios of filament widths for different tracers could be a useful tool to estimate the accretion rate onto the filament.
\end{abstract}


%%%%%%%%%%%%%%%%%%%%%%%%%%%%%%%%%%%%%%%%%%%%%%%%%%

%%%%%%%%%%%%%%%%% BODY OF PAPER %%%%%%%%%%%%%%%%%%

\section{Introduction}
\label{sec:intro}

%{\bf
The filamentary nature of molecular clouds has been clearly demonstrated in the astronomical literature, specially through observations of the thermal emission from dust \citep[][among others]{Andre+14,Molinari+10,Wang+15,Hacar+18,PlanckXXXII}.
The filaments embedded in the clouds are frequently organized in hub-filament systems.
Longitudinal gas flow directs material to the hubs sitting at the intersection of two or more filaments and, while it has been suggested that these hubs are the sites of high-mass star formation, it is a matter of debate
\citep{Peretto2013,Henshaw2014,Yuan+18,Lu+18,Dewangan2020}.
At the same time, molecular filaments are observed to contain cores, usually as result of gravitational fragmentation within the filament.
Star formation within these embedded cores constitutes a non-trivial source of low- and intermediate-mass stars
\citep{Andre+14,Tafalla+15}.
Although the internal density and velocity structure of molecular filaments would help us understand the mass reservoir available for star formation, the details of this mass flow and its impact on the star formation phenomenon is still an open question in the community.
%}
Different models have been proposed for their internal structure, but observations that discern between those
models are elusive.

The simplest models assume that molecular filaments are structures in hydrostatic equilibrium
\citep{Sto63,Ostriker64,InutsukaMiyama92,FischeraMartin12,Burge+16}. However, these do not include a mechanism
for the origin for the filamentary structure.
In contrast, another set of models state that the filaments are formed where turbulent flows within the cloud
meet \citep[e.g.][]{Padoan+01,Auddy+16},
although this convergence of flows may not be a consequence of turbulence but of the global gravitational collapse 
of the cloud \citep{Heitsch13a,Heitsch13b,HennebelleAndre13,Zamora+17}.

In a spherically-symmetric collapse, the mass that flows across a constant-radius surface has to be conserved,
and so this flux equates to the time derivative of the mass within the given surface.
In contrast, in the case of a collapsing filament, the rate of change of mass within a given \emph{cylindrical}
radius does not have to be equal to the mass flux across that radius since the mass may be evacuated in the
longitudinal direction, in a similar fashion to a squeezed toothpaste tube.
So, a filament formed due to the global collapse of a cloud acts as a funnel that redirects the gas around the cloud
towards star-forming clumps located at the hubs where filaments meet, or to the smaller mass clumps formed
within the filaments themselves.
In this picture, proposed by \citet[GV14 from here on]{GV14}, the molecular filaments are not material but river-like
flow structures along which the gas falls down the large-scale gravitational well.
Note that this model relies on gravitational collapse only and
does not require external agents to direct the gas to flow along the filaments \citep[for example,][]{Balsara+01}.

Longitudinal flows consistent with the scenario described above { have been widely observed in a number of filamentary molecular clouds such as Taurus \citep{Dobashi+19}, Serpens \citep{Kirk+13, Lee+14, FernandezLopez+14, Gong+18}, Musca \citep{Bonne+20}, DR21 \citep{Hu+21}, clouds of the Galactic Center \citep[SgrB2(N),][]{Schworer+19}, or filamentary high-mass star-forming regions \citep[e.g.,][]{Peretto+14, Tackenberg+14, Lu+18, Veena+18, Yuan+18, Chen+19, Issac+19, TrevinoMorales+19, Chung+21, Liu+21, Ren+21}. 
In addition, possible signs of accretion onto filaments have also been reported in the literature \citep[e.g.,][]{Gong+18, Shimajiri+19, Chen+20a, Chen+20b, Sepulveda+20, Zhang+20, Gong+21}.
Furthermore,} magnetic structures derived from the longitudinal flow might be used to determine the physical conditions of the gas \citep{Gomez+18}. { A few recent works have reported the detection of a change in the magnetic field orientation from perpendicular to the filament outside the filament to parallel inside it \citep{Qiu+13, Pillai+20, Arzoumanian+21, Palau+21}. All these observed properties (velocity gradients along the filaments, accretion onto the filaments, and change in the magnetic field orientation from perpendicular to parallel to the filament) are consistent with the results of the simulations of GV14 and \citet{Gomez+18}.}

\begin{figure}
\centering
\includegraphics[width=0.50\textwidth]{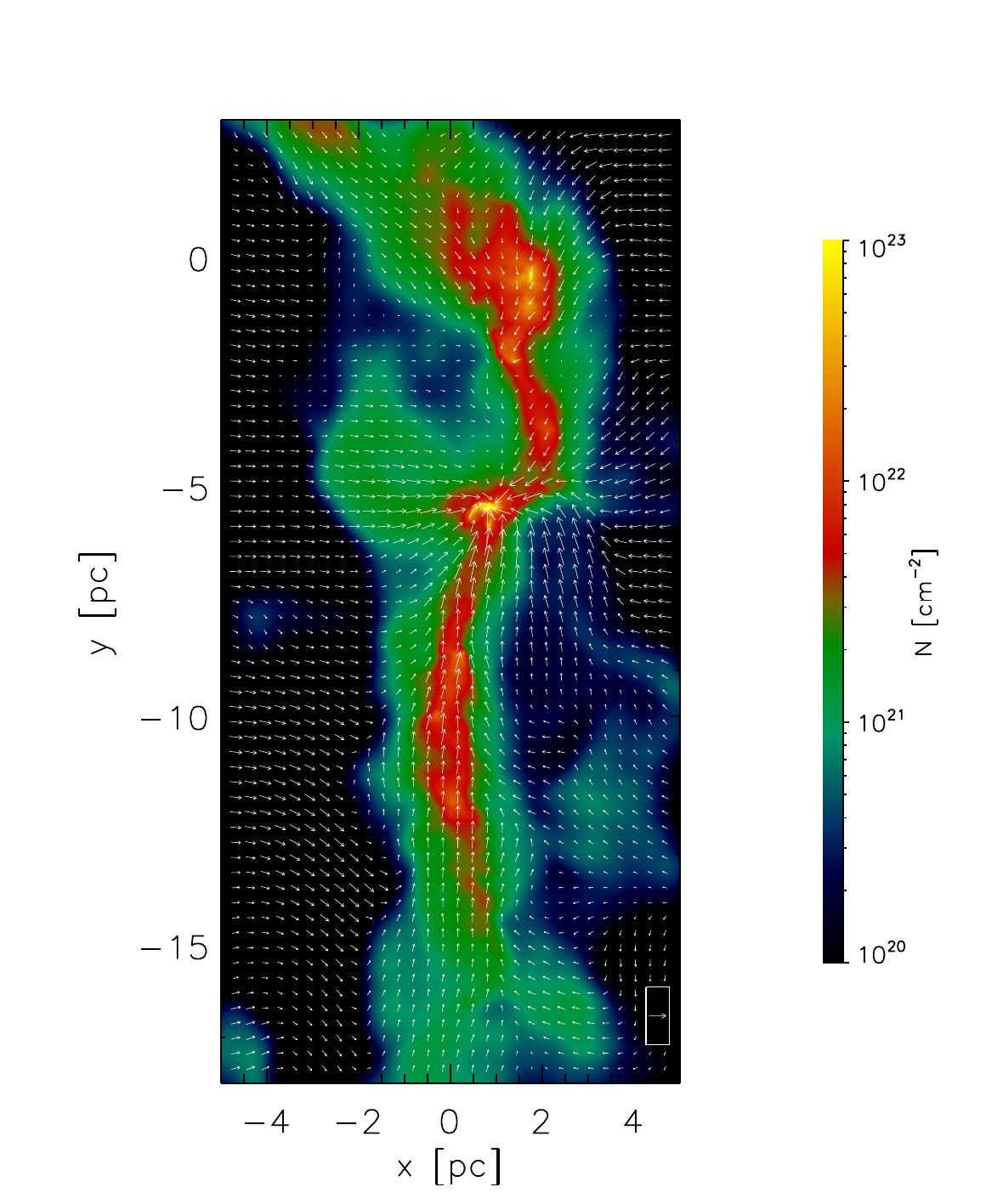}
\caption{Filament 2 from GV14 (section of their fig. 3), at $26.56 \Myr$ into their simulation.
Colours show the column density of the gas, while the arrows show the density-weighted projected velocity,
with the arrow in the lower right representing $2\kms$.}
\label{fig:filament}
\end{figure}

The `filaments as flow structures' model naturally arises from the global hierarchical collapse model
\citep[GHC,][]{Vazquez+19},
in which the clouds are pictured as containing collapsing regions within larger-scale collapses,
each scale accreting from larger ones.
As the warm neutral medium experiences a compression and undergoes a phase transition to cold neutral medium
\citep{HennebellePerault99,KoyamaInutsuka02},
its density increases by a factor of $\sim 100$, while its temperature decreases by a similar factor.
As a consequence, the Jeans mass in the recently formed cloud drops by a factor of $\sim 10^4$,
and so the collapse proceeds in an almost pressureless fashion.
Filaments formed in this scenario arise from the anisotropic collapse of the gravitationally unstable structure,
in which a triaxial spheroid collapses along its shortest dimension first, thus sequentially forming sheets and
filaments \citep{Lin+65}.
In this sense, the filaments are the actual collapse flow from the large to small scales:
since the molecular cloud is formed by a compression (and because of its near-pressureless subsequent collapse),
it is expected to have a sheet-like geometry, which fragments and further collapses into filaments.
The filaments will accrete gas in a direction approximately perpendicular to them and redirect the flow to their
longitudinal direction, thus evacuating the filament.
This flow around and along the filaments is shown in Figure \ref{fig:filament}, reproduced from GV14.
Those authors noticed that the above mentioned flow redirection happens smoothly, without shocks.
This is possible, again, by the longitudinal evacuation of the accreted gas since a pressure excess simply increases
the longitudinal flow rate.
Additionally, those authors noticed that the filaments in their simulations appear to reach a steady state, so that they
exist for times longer than their flow-crossing time.

{ Verification of this mode of filament structure and evolution may be achievable by exploiting the 
tools available in astrochemistry.  
Molecular line observations of known tracers of molecular cloud structure at sufficient sensitivity and spatial and spectral resolution can reveal various properties including 
the %{\bf
density and temperature \citep[e.g.,][]{Busquet+13,Hacar+13,Feher+16,Xiong+17,Liu+18,Sepulveda+20,Schmiedeke+21}, and velocity flow (or accretion rate) into and through the filament \citep[e.g.,][]{Lee+14,Dhabal+18,Guo+21,Zhou+21}.}
To probe the accretion rate into the filament, one can exploit the chemical timescales of molecule formation from the diffuse to the dense interstellar medium.%}
In contrast to an equilibrium state,
a gas parcel falling into a filament experiences a changing physical environment.
So, a given molecular species will trace the density distribution only if the molecule is
able to form faster than the timescale at which the density and temperature changes.
The abundance of a slow-forming molecule, in contrast, will lag behind the actual gas density of the parcel
it is observed in.
As a consequence, a filament observed using a fast-forming molecule should appear wider than the same filament
observed using a slow-forming molecule.
This change in the filament thickness when measured with different molecular lines should reflect the
accretion rate that the filament experiences.

In this contribution, we explore the % molecular abundances formed out of equilibrium 
time-dependent molecular abundances
in the accretion flow
feeding a filament and contrast them with the abundances expected in a steady-state distribution.
In \S \ref{sec:model} we review the gas distribution model and the chemical network used.
In \S \ref{sec:abundances} we explore the resulting abundance distributions for a subset of the molecular
species calculated, and explore the effect of changing the flow of mass towards the filament.
Finally, in \S \ref{sec:summary} we present a discussion of the results.

%%%%%%%%%%%%%%%%%%%%%%%%%%%%%%%%%%%%%%%%%%%%%%%%%%
\section{Filament model}
\label{sec:model}

%%%%%%%%%%%%%%%%%%%%%%%%%%%%%%%%%%%%%%%%%%%%%%%%%%
\subsection{{ Physical model}}

%%%%%%%%%%%%%%%%%%%%%%%%%%%%%%%%%%%%%%%%%%%%%%%%%%
\subsubsection{{ Accreting, }hydrodynamical model}
\label{sec:hydro}

GV14 analyzed two gaseous filaments formed in a simulation similar to one presented in \citet{Vazquez+07}.
They used the SPH code {\sc gadget-2} to follow the evolution of two colliding streams of gas at a density and temperature
similar to the warm neutral medium.
The code was modified to include the cooling function proposed by \citet{KoyamaInutsuka02}
(with typographical errors corrected as outlined in \citealt{Vazquez+07}), so that the gas is thermally bistable.
The streams move initially in opposite directions { towards each other} at $1.2~c_s$, where $c_s$ is the adiabatic sound speed, inducing a
phase transition to the cold neutral medium at the region where the flows collide.
This dense layer accretes gas from the remnants of the initial flows, thus increasing its mass and ultimately reaching
a density larger than $5 \times 10^{4}~\mathrm{M_{\odot}}$.
The resulting dense cloud is ram-pressure confined by the inflowing gas and it is therefore subjected to
hydrodynamical instabilities \citep{Vishniac94,WalderFollini00,Heitsch+05,Heitsch+06,Vazquez+06}.
These instabilities create moderately supersonic turbulence within it and, in turn, induce non-linear density fluctuations.
Although the layer's mass increases in time and eventually becomes gravitationally unstable, the individual
density enhancements collapse before the cloud as a whole.
An important advantage of such a simulation setup is that the internal turbulence is generated in a self-consistent
way, and so we do not need to impose a driving rate. Imposing this would carry the risk of over-driving the
turbulent flow and artificially supporting the cloud against a global gravitational collapse.
Additionally, the induced density fluctuations and \emph{remnant} surrounding flow will also be consistent with one
another.
We refer the reader to \citet{Vazquez+07} for further details on the global evolution of the molecular cloud.

The filaments formed in this simulation are long-lived, being clearly discernible as long as there is enough
gas falling into them.
They appear to reach a steady state since the accreted gas is evacuated along the filament onto clumps formed
within the filament or at the position where two or more filaments meet, thus forming a hub-and-spoke distribution.
GV14 report filament lengths of $\sim 15\pc$, masses of $\sim 10^3~{\mathrm{M}_\odot}$ (depending on the density threshold used
to define the structures), a flattened { density} profile with a core size of $\sim 0.3\pc$ and a power-law envelope with
an approximately $-2.5$ logarithmic slope.
The measured accretion rate toward the filament is $10~{\mathrm{M}_\odot}~\pc^{-1}~\Myr^{-1}$ (also depending on the density
threshold used to define it).

\begin{figure}
\centering
\includegraphics[width=0.49\textwidth]{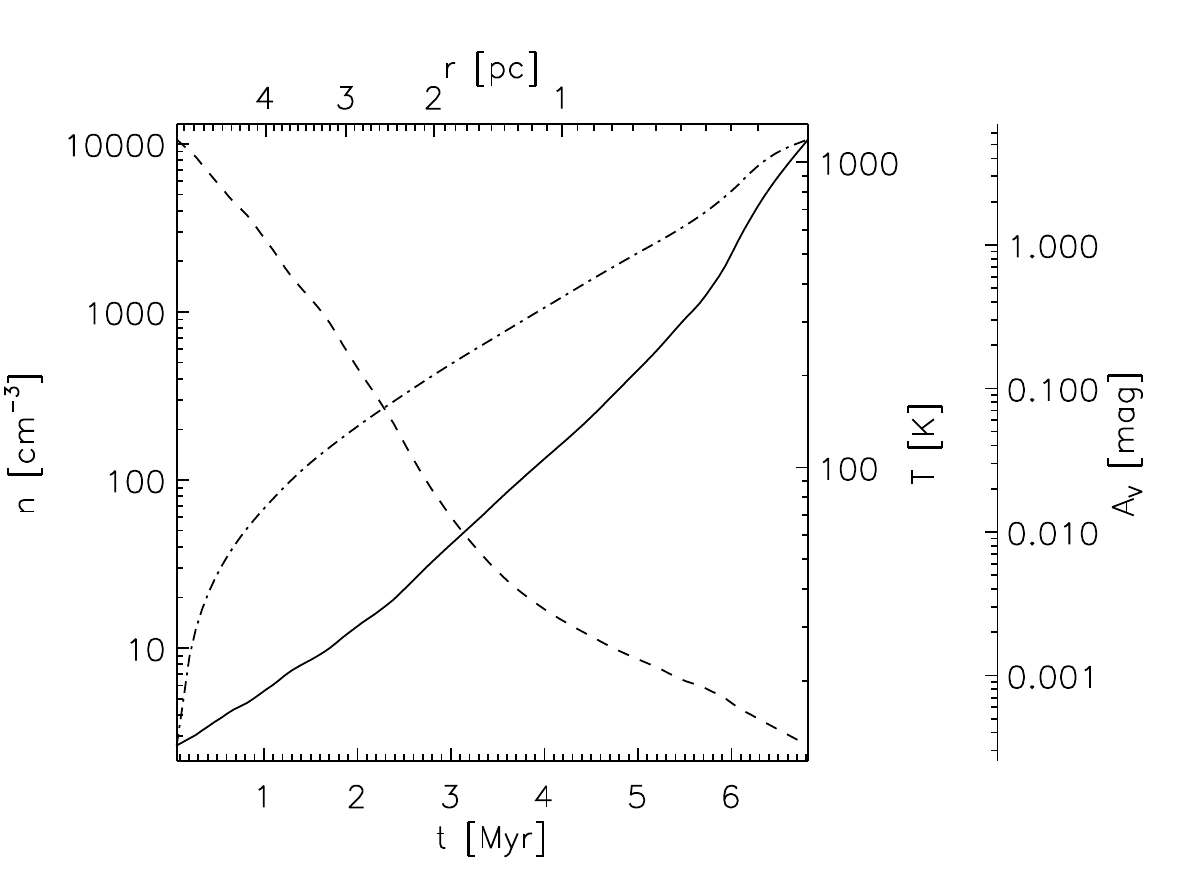}
\caption{Density ({\it solid}), temperature ({\it dashed}), and visual extinction ({\it dash-dotted})
evolution of a gas parcel as it is accreted into the filament for the nominal accretion case (see \S\ref{sec:abundances}).
The physical conditions that the parcel experiences are plotted as a function of time and distance from the center of the 
filament.}
\label{fig:density_temp}
\end{figure}

While being accreted, a gas parcel will experience an increasing density, decreasing temperature,
and increasing visual extinction.
By assuming that the filament is in steady state, we can use the azimuthally and longitudinally averaged radial
velocity from GV14 to obtain the %density and temperature
physical conditions
a gas parcel experiences as a function of time (see Fig. \ref{fig:density_temp}).
In this case, the density increases from $\sim 3$ to $\sim 10^4 \pcc$ in an approximately exponential
form (with an $\mathrm{e}$-folding time of about $0.83 \Myr$), the { gas} temperature decreases from $\sim 10^3$ to $\sim 10 \degK$
and the visual extinction reaches a value of $\sim 5.4 \,\mathrm{mag}$ { at the centre of the filament}.

%%%%%%%%%%%%%%%%%%%%%%%%%%%%%%%%%%%%%%%%%%%%%%%%%%
\subsubsection{Non-accreting, static model}
\label{sec:static}

We consider two scenarios in this work: i) one in which we compute the chemical evolution of the parcel as it infalls into the filament, and ii) another in which we compute the chemical evolution as a function of time only at fixed positions along the flow (i.e., we decouple the chemistry from the flow).  
For this latter case, we consider the density and temperature distribution to be static in time, i.e.,
the filament does not accrete from its surroundings.
Here, each gas parcel experiences the physical conditions shown in Figure \ref{fig:density_temp}
as a function of radius only (without the time dependence in the bottom axis).
We note here that the chemistry is fully decoupled from the hydrodynamics, i.e., the chemical structure is calculated by post-processing the simulated physical conditions resulting from the hydrodynamics simulations. We later discuss the implications of this assumption.

%%%%%%%%%%%%%%%%%%%%%%%%%%%%%%%%%%%%%%%%%%%%%%%%%%
\subsection{Chemical model}
\label{sec:chemistry}

The chemical model used in this work is described in detail in \citet{Walsh12} and \citet{Walsh14} and we provide a brief description here.  
This model was originally developed for use in protoplanetary disks but is inherently flexible, and has been used in a variety of environments including the envelopes of 
forming protostars \citep[e.g.,][]{Drozdovskaya14}, shocks in outflow cavity walls \citep[e.g., ][]{Palau17}, dark clouds \citep[e.g.,][]{Penteado17}, and the reaction network has also been used in models of circumstellar outflows \citep[e.g.,][]{Vandesande21}.

%{\bf 
Because we are simulating the chemical evolution from the diffuse to the dense interstellar medium, it is necessary to include all competing chemical processes that lead to the key molecular tracers considered in this work which range from simple diatomics (e.g., CO and CN) to larger organic molecules (e.g., \ce{H2CO} and \ce{CH3OH}). For example, as the visual extinction and the density increase and the temperature decreases, freeze-out onto dust grains of abundant elements such as oxygen, begins to compete with gas-phase formation of simple molecules such as CO. 
The freeze-out of oxygen atoms eventually leads to the formation of water ice via grain-surface hydrogen-addition reactions \citep[see, e.g.,][]{vandishoeck2013}. 
As the temperature decreases further ($\lesssim 25$~K) CO begins to also freeze-out which triggers the formation of larger molecules via atom-addition reactions forming molecules such as methanol \citep[\ce{CH3OH}; see, e.g.,][]{linnartz2015}.  
At low temperatures, molecules can be returned to the gas phase by non-thermal mechanisms such as photodesorption \citep[see, e.g.,][and references therein]{Cuppen17}; hence, the inclusion of ice formation and ice chemistry also requires the need to include all possible mechanisms for ice destruction. 
As explained in the review of molecular cloud chemistry by \citet{agundez2013}, such large chemical networks are highly non-linear and stiff and specialised solvers are required (e.g., ODEPACK\footnote{\url{https://computing.llnl.gov/projects/odepack/software}}). 
This is also the motivation for decoupling the chemistry from the physics and computing the chemistry in a post-processing manner. 
It remains computationally demanding to couple comprehensive chemistry networks with hydrodynamics simulation with modern attempts limited to reduced networks simulating the chemistry of simple species only \citep[see, e.g.,][for the case study of protoplanetary disks but the discussion of which is generally applicable]{haworth2016}.%}

The model includes gas-phase chemistry based on the Rate12 version of the UMIST Database for Astrochemistry (UDfA\footnote{\url{http://www.udfa.net}}; \citealt{McElroy13}), 
supplemented with gas-grain chemistry derived from the OSU 2008 chemical network \citep{Garrod08,Laas11}. 
We include self- and mutual-shielding of \ce{H2}, CO, and \ce{N2} to photodissociation by interstellar photons using the shielding functions from \citet{Heays17}. 
The gas-grain processes included are freezeout onto dust grains, and desorption (sublimation) into the gas-phase via 
thermal desorption, photodesorption by both external UV photons and by secondary UV photons induced by the interaction of cosmic rays with molecular hydrogen, 
and reactive desorption.
The binding energies needed to calculate the thermal desorption rates (and grain-surface reaction rates) are those compiled for use with UDfA. 
We used laboratory-measured photodesorption rates where available. 
We also include grain-surface reactions to allow the formation and processing of ices on dust grains.  
These reactions include atom-addition reactions, radical-radical recombination, and photodissociation of ices. 
The reaction rates for gas-grain processes are calculated according to the prescriptions recommended in \citet{Cuppen17}.
We adopt a two-phase model in this work, i.e., we do not discriminate between chemistry occurring in the 
bulk ice and that occurring on the surface of the ice mantles. 
However, we do restrict that active grain-surface chemistry occurs within two monolayers only in the ice mantle. 
We include the formation of molecular hydrogen via the grain-surface recombination of atomic hydrogen.

%{\bf 
Because the filament evolution starts with a high temperature, low density, and low A$_\mathrm{v}$, i.e., conditions representative of the diffuse 
interstellar medium, we begin the chemistry with atomic initial conditions using the abundances presented in Table~\ref{tab:abundances} and that are reproduced from \citet{McElroy13}, the only 
difference being that we assume here that all hydrogen is initially in atomic form.  
The abundances correspond to the values measured for the elemental composition of the local diffuse 
interstellar medium in the Milky Way and account for depletion of heavy elements (i.e, metals) onto to refractory dust grains \citep{Graedel82}.
The chemical evolution of the filament is calculated over the averaged physical conditions presented in Fig.~\ref{fig:density_temp}. %}
In this way we follow the formation of molecules along a typical parcel of gas falling into the filament that experiences an increasing density and A$_\mathrm{v}$ and a decreasing temperature. 
Note that most molecular formation under these conditions occurs via gas-phase chemistry; gas-grain and grain-surface processes only begin to become 
important once the A$_\mathrm{v}$ increases to $\gtrsim$~3~mag, and the temperature falls below $\sim$~100~K, at which point atoms like atomic oxygen begin 
to stick to dust grains and the formation of water ice begins as described above.  

One caveat in our adopted methodology is that we assume that the dust and gas temperatures are equal along the flow.  
In reality these temperatures will decouple at low density in irradiated clouds due to the competition between gas heating and cooling \citep[see, e.g.,][]{Hollenbach99}.
Hence, our simulations are not self-consistent in that the physics and chemistry of the filament are decoupled; however, this assumption is 
necessary in order to include the level of chemical complexity needed to simulate all possible observable molecular tracers considered in this work.

%We consider two scenarios in this work: i) one in which we compute the chemical evolution of the parcel as it infalls into the filament, and ii) another in which we compute the chemical evolution as a function of time only at fixed positions along the flow (i.e., we decouple the chemistry from the flow).  
%We run this latter case
{ For the nominal, accreting filament scenario, the chemical evolution is computed in the changing physical conditions shown in Figure \ref{fig:density_temp}.
For the non-accreting case, the chemical evolution is calculated at fixed conditions}
for a time of 63~Myr which should be sufficiently long for the gas-phase abundances to reach steady state across the filament.
We do this to test if the formation timescale of molecules can be used as a diagnostic of the flow rate into the filament. 

\begin{table}
    \begin{center}
        \caption{Initial elemental abundances (with respect to total H nuclei) from \citet{McElroy13}. }
        \label{tab:abundances}
        \begin{tabular}{lc}
            \hline
%            Species & $ $ \\
            Species & Fractional abundance \\
             \hline
             H     &     1.00  \\
             He    &     $9.75\times 10^{-2}$  \\
             O     &     $3.20\times 10^{-4}$\\
             C     &     $1.40\times 10^{-4}$ \\
             N     &     $7.50\times 10^{-5}$ \\
             S     &     $8.00\times 10^{-8}$ \\
             F     &     $2.00\times 10^{-8}$ \\
             Si    &     $8.00\times 10^{-9}$ \\
             Mg    &     $7.00\times 10^{-9}$ \\
             Cl    &     $4.00\times 10^{-9}$ \\
             Fe    &     $3.00\times 10^{-9}$ \\
             P     &     $3.00\times 10^{-9}$ \\
             Na    &     $2.00\times 10^{-9}$ \\
            \hline
        \end{tabular}
    \end{center}
\end{table}

%  349 F          2.000e-08   19.0
%  474 Cl         4.000e-09   35.0
%  569 Fe         3.000e-09   56.0
%  570 Mg         7.000e-09   24.0
%  573 Na         2.000e-09   23.0
%  592 P          3.000e-09   31.0
%  650 Si         8.000e-09   28.0
%  670 S          8.000e-08   32.0
%  680 He         9.750e-02    4.0
%  681 N          7.500e-05   14.0
%  684 O          3.200e-04   16.0
%  685 C          1.400e-04   12.0
%  689 H          1.000e+00    1.0
%%%%%%%%%%%%%%%%%%%%%%%%%%%%%%%%%%%%%%%%%%%%%%%%%%
\section{Abundances}
\label{sec:abundances}

\begin{figure*}
\centering
\includegraphics[width=\textwidth]{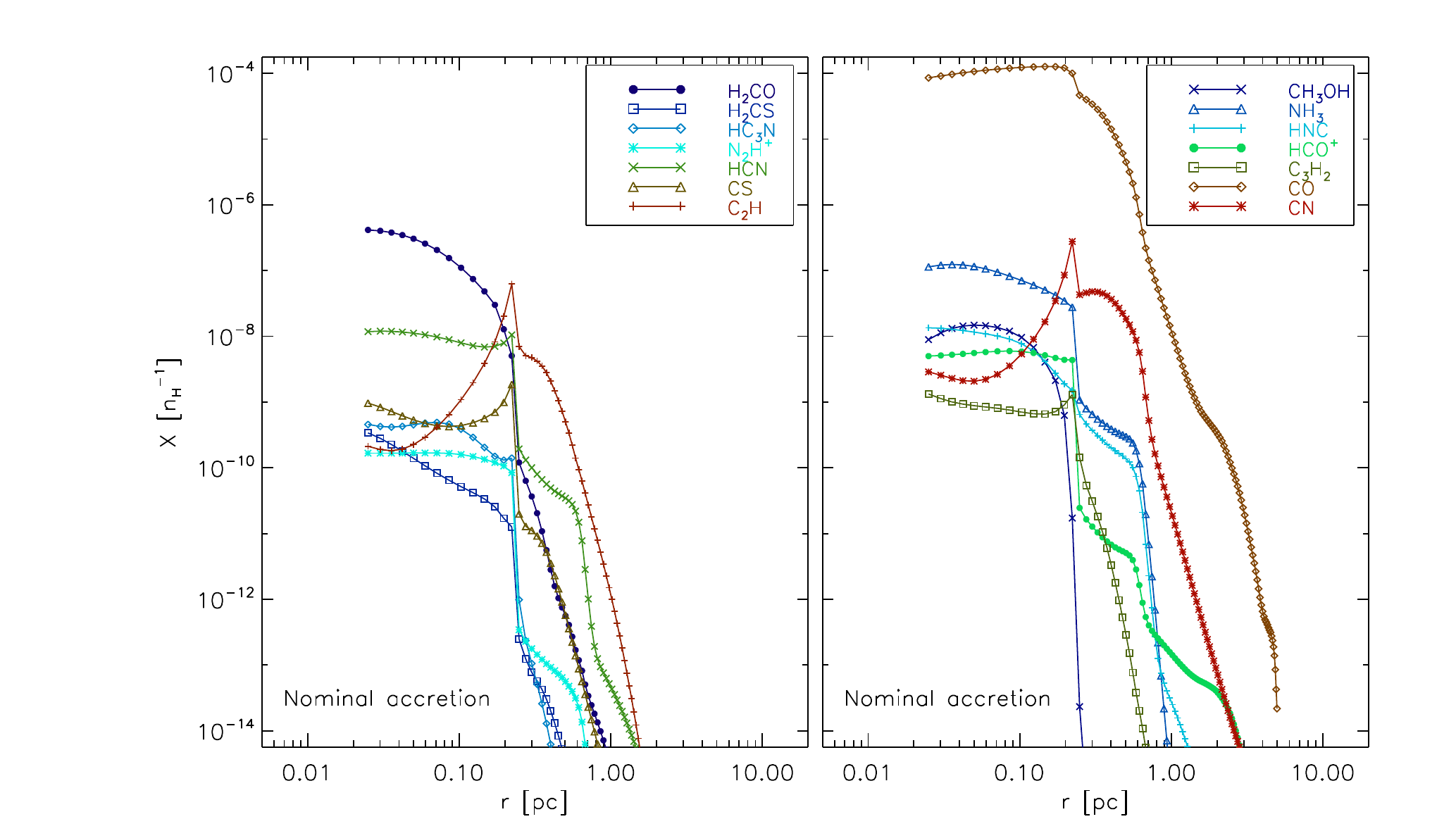} \\
\includegraphics[width=\textwidth]{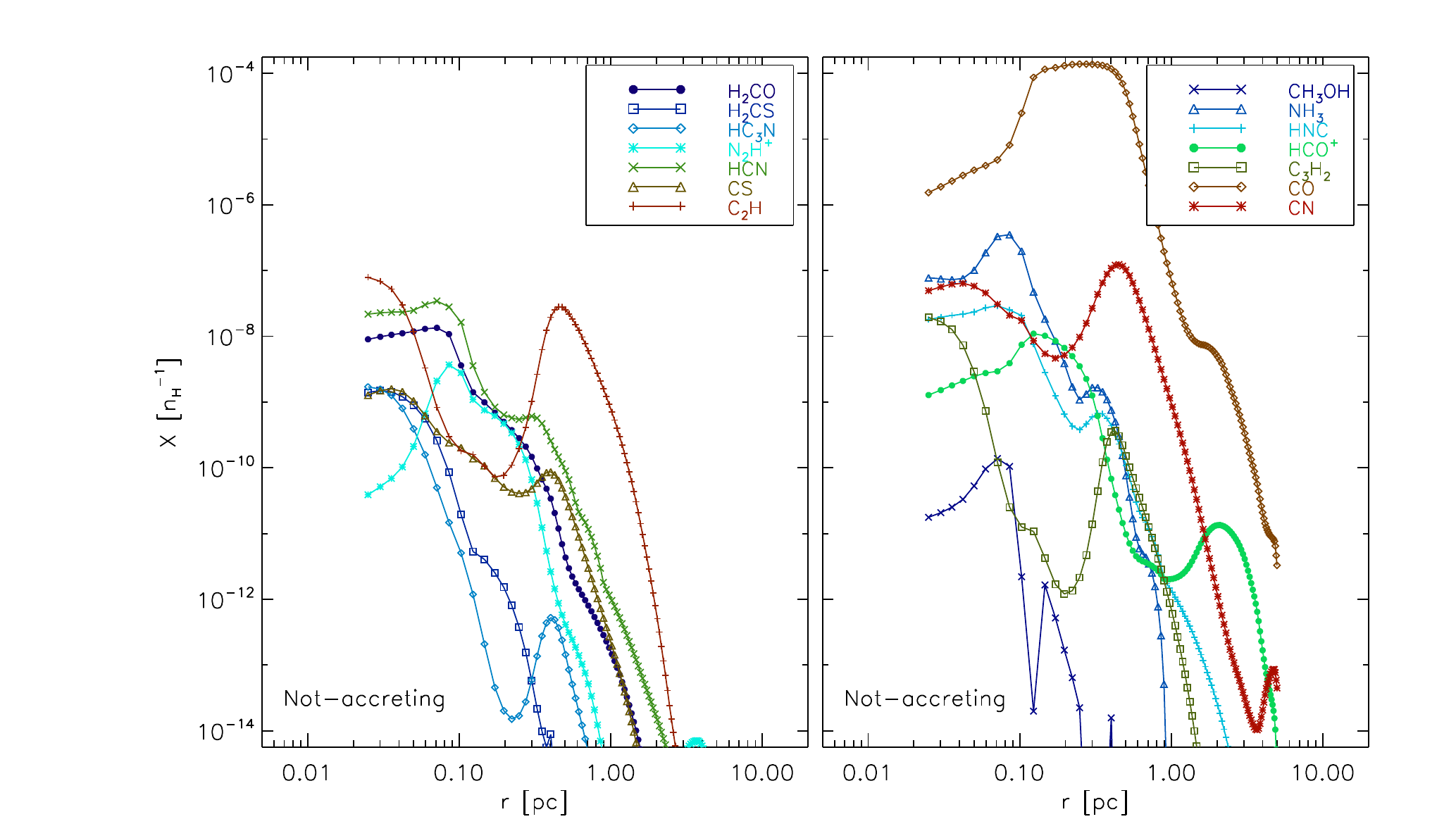}
\caption{Time dependent (i.e. accreting case, {\it top row})
and steady state molecular abundances (non-accreting case, {\it bottom row})
for a set of molecular species.
Differences between both abundance distributions reflect the changing density and temperature environment
for molecules with slow chemical timescales.}
\label{fig:abun}
\end{figure*}

A molecule with a short chemical timescale, i.e. a fast-forming molecule, will reach its steady-state
abundance before the gas parcel's environment changes significantly.
In this case, the given molecular abundance will accurately reflect the local physical conditions within the filament.
In contrast, a molecule with a long chemical timescale, i.e. a slow-forming molecule,
will not reach its steady-state abundance before the gas parcel
falls more deeply into the filament and its environmental conditions change.
In this case, the slow-forming molecule abundance \emph{will not} reflect the filament density and temperature.
Therefore, steady-state and time-dependent abundances will differ for slow-forming molecules, although the
interdependencies of the formation rates of the molecular species means that the abundance of fast-forming molecules
will also differ.
Both cases are shown in Figure \ref{fig:abun}.

{ The top panels of Fig.~\ref{fig:abun} show the fractional abundances of several common and observable molecular tracers of cold and dense gas for the case in which the chemistry onto the filament is calculated along the accretion flow (i.e., the chemistry and flow are coupled). 
The distance along the flow at which a molecule reaches an appreciable gas-phase fractional abundance ($\gtrsim 10^{-11}$) varies between $\sim 1 - 2$~pc for 
molecules such as CO, CN, \ce{C2H} (formed the farthest out), HCN, HNC, and \ce{NH3}, and $\sim 0.2 - 0.9$~pc for other species considered here including larger molecules such as \ce{H2CO}, \ce{H2CS}, \ce{HC3N}, and \ce{CH3OH} (formed the farthest in). 
This structure mimics to some extent the chemical structure predicted by chemical models of weak to moderate photon-dominated regions \citep[see, e.g., the recent models of the chemical structure of the Horsehead nebula by][]{LeGal17}. 
In these models, radicals such as CN and \ce{C2H} sit closer to the edge of the cloud than chemically-related molecules such as HCN and \ce{c-C3H2}, respectively. 
Large molecules such as \ce{CH3OH} take longer to form along the flow due to their reliance on grain-surface chemistry.  
For instance, \ce{CH3OH} can only form after the gas-phase formation of CO.  
Once the temperature is sufficiently low, CO can begin to stick to grain surfaces where it is hydrogenated to form \ce{H2CO} and \ce{CH3OH}.  
\ce{H2CO} and \ce{CH3OH} are then returned to the gas-phase via non-thermal desorption, in this case, via a combination of photodesorption and reactive desorption.}

{ The bottom panels of Fig.~\ref{fig:abun} show the fractional abundances of the same molecules for the case where the chemistry is computed in time only, i.e., the chemistry and flow are decoupled. 
The general pattern described above still persists; however, there are some differences.  
The larger molecules (\ce{HC3N}, \ce{H2CS}, and \ce{CH3OH}) now only reach appreciable abundances ($\gtrsim 10^{-11}$) much farther into the filament ($\sim 0.1$~pc). 
As the chemistry here is close to steady-state, the results show that the destruction timescale of these species at distances beyond $\sim 0.1$~pc are shorter than 
63~Myr. 
The most profound difference is in the composition of the gas entering the densest point of the filament.  
Several species have fractional abundances that differ by more than an order of magnitude between the two cases.  
%The fractional abundance of CN and \ce{C2H} are increased in the static case compared with the accreting case by one and two orders of magnitude, respectively.  
The CN and \ce{C2H} fractional abundances are higher in the static case compared with the accreting case by one and two orders of magnitude, respectively.  
This is due to the long timescales required for cosmic-ray reactions to occur in dense gas, which can ionise helium, which goes on to dissociate and ionise molecules, creating radicals. 
%Further, the abundances of both \ce{H2CO} and \ce{CH3OH} are decreased in the static case by around two orders of magnitude at the same point.  
Further, the \ce{H2CO} and \ce{CH3OH} fractional abundances are lower in the static case by around two orders of magnitude at the same point.  
This is again due to the long timescales needed to destroy molecules via cosmic-ray reactions. 
Whilst in the coupled case, most of the fractional abundances increase monotonically along the flow (with the exception of the radicals CN and \ce{C2H}), the abundances in the static case show more complex morphologies along the filament, with many species having a double-peaked abundance structure. 
For instance, \ce{C2H} exhibits a peak in abundance at $\approx 0.5$~pc in the static case, that is mirrored by the chemically-related species, \ce{c-C3H2}, 
and there is a similar behaviour in the case of CN, HCN, HNC, and \ce{HC3N}.
}

{ The abundance behaviour described above along with the results shown in Fig.~\ref{fig:abun} show that the formation (and destruction) 
timescale of all considered molecules are indeed longer than the time spent at each location along the flow.

That is, we have demonstrated that the chemistry along the flow is not at steady state.}
Hence, the accretion towards the filament implies that slow-forming molecules will not reach their chemical timescale until
later times, corresponding to regions closer to the filament crest, and in the accreting case we should expect, as a first
approximation, narrower distributions for slow-forming molecules compared to the distributions of fast-forming molecules.

\begin{figure*}
\centering
\includegraphics[width=\textwidth]{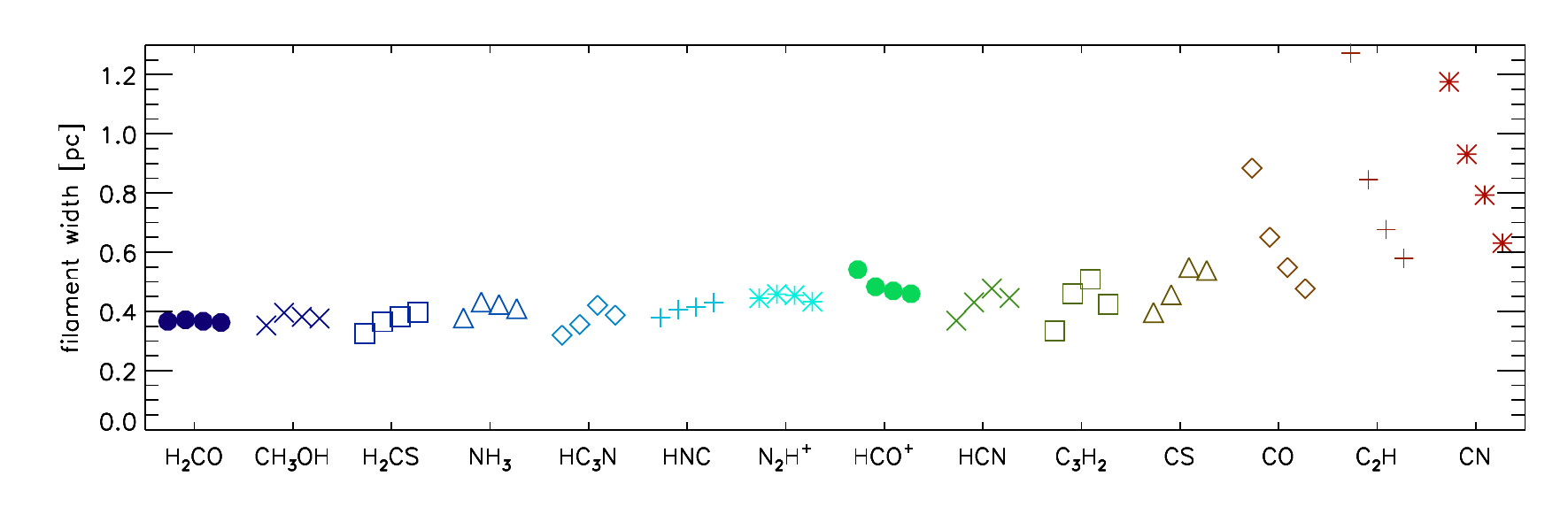}
\caption{Filament width measured as the full-width at half maximum for the convolved, projected molecular abundance distributions.
Data points for each molecular species correspond to, from left to right, the steady state, half the accretion rate,
nominal accretion, and twice the accretion rate cases.
Colours and symbols are the same as in Fig. \ref{fig:abun}.}
\label{fig:fwhm}
\end{figure*}

The complicated distribution of abundances makes the width at a given threshold abundance an unreliable measure of the
accretion effect on the molecular chemistry.
A simple analytical function fitting would also be of little use.
Instead, we measured the abundance widths using the full-width at half maximum (fwhm) of the projected column density
distribution convolved with a Gaussian profile with $0.1\pc$ size, as this would better reflect an astronomical
observation.
{ The resulting smoothed abundance distributions are available as a supplementary online figure} and the corresponding
width values for selected molecules are shown in Figure \ref{fig:fwhm}.

An important aspect of the adopted model filament is the redirection of the accretion toward the filament
to longitudinal flow (see \S\ref{sec:intro}).
This longitudinal evacuation of gas adds a degree of freedom to the conservation of mass, namely the
mass flux along the filament.
%For a filament in steady state, the continuity equation in cylindrical coordinates reads,
%
%\begin{equation}
%    \frac{1}{r}\frac{\partial}{\partial r}\left(r \rho v_r\right)
%      + \frac{\partial}{\partial z}\left(\rho v_z\right) = 0.
%\end{equation}
%
%\noindent
%Thus,
Since the continuity equation in steady state is linear in the velocity, we may
multiply the radial and longitudinal velocities by a constant factor and retain the same density profile.
Although a fully self-consistent model for the density and velocity structure of the filament is beyond the scope
of this contribution (Gómez \& Vázquez-Semadeni, {\it in prep.}), here we increase (or decrease) the radial velocity
of the flow in order to emulate different accretion rates.
These new radial velocities will imply that a gas parcel experiences a more rapidly (or more slowly) changing environment, yielding different molecular abundance profiles.

GV14 reported an accretion rate to the filament of $10~{\mathrm{M}_\odot} \pc^{-1} \Myr^{-1}$.
Together with the static model and this nominal accretion rate, Fig.~\ref{fig:fwhm} shows the calculated fwhm
values for cases with double and half the nominal accretion rate.
Widths measured for the \ce{C2H}, CN, and CO molecules are the most sensitive to the changing accretion toward the
filament, with thinner distributions as the accretion rate increases.
On the other hand, widths for some other molecules like N$_2$H$^+$ and H$_2$CO, remain almost constant for different
accretion rates.
Yet other width values, like those for CS, HNC, and H$_2$CS, follow the opposite trend, with the
molecule distribution growing wider with
increasing accretion rate, although only marginally (for the density profile used here).
For other molecules, like C$_3$H$_2$ and HCN, the corresponding width values show a non-monotonic dependence on
accretion rate.

%%%%%%%%%%%%%%%%%%%%%%%%%%%%%%%%%%%%%%%%%%%%%%%%%%
\section{Discussion}
\label{sec:summary}

\subsection{Molecule pairs tracing accretion onto filaments}

In the GHC scenario, filaments represent the locus of gas accretion from the cloud to the star-forming clumps.
Therefore, the filaments are not density but flow structures as the gas within them is constantly replenished.
This means that a gas parcel within the filament will experience changing densities, temperatures, and opacities,
at a rate depending on the accretion rate toward the filament.
Since chemical abundances depend on these physical conditions, molecular distributions should reflect the accretion.
Furthermore, the chemical timescales will be convolved with the accretion timescale, so that faster-forming molecules
will better reflect the \emph{local} physical conditions a gas parcel experiences than slower-forming ones,
which will retain memory of the density and temperature external to the actual position of the gas parcel.

\begin{table}
    \begin{center}
        \caption{Molecular species ordered by increasing sensitivity to accretion. %,
        %measured as the absolute value of the logarithm of the 
        %ratio of widths measured for the nominal accretion to the static models.
        }
        \label{tab:sensitivity}
        \begin{tabular}{llcc}
            \hline
%            Molecular species & $ \log( w_{\mbox{nominal}}/w_{\mbox{static}} ) $ \\
            & Molecule & $ w_{\mbox{nominal}} $ & $ w_{\mbox{nominal}}/w_{\mbox{static}} $ \\
            &          &             [pc]       &                              \\
            \hline
            \multirow{6}*{Insensitive to accretion} 
                                     & H$_2$CO      &     0.37 &    1.00 \\
                                     & N$_2$H$^+$   &     0.45 &    1.02 \\
                                     & CH$_3$OH     &     0.38 &    1.08 \\
                                     & HNC          &     0.41 &    1.09 \\
                                     & NH$_3$       &     0.42 &    1.12 \\
                                     & HCO$^+$      &     0.47 &    0.87 \\
            \hline
            \multirow{3}*{Marginally sensitive}     
                                     & H$_2$CS      &     0.38 &    1.17 \\
                                     & HCN          &     0.48 &    1.29 \\
                                     & HC$_3$N      &     0.42 &    1.32 \\
            \hline
            \multirow{5}*{Sensitive to accretion}   
                                     & CS           &     0.55 &    1.38 \\
                                     & CN           &     0.79 &    0.67 \\
                                     & C$_3$H$_2$   &     0.51 &    1.52 \\
                                     & CO           &     0.55 &    0.62 \\
                                     & C$_2$H       &     0.68 &    0.53 \\
            \hline
        \end{tabular}
    \end{center}
\end{table}

In this contribution, we propose that the widths measured using different molecular species can be used to measure the accretion toward the filament.
In our Figure \ref{fig:fwhm}, we show that the distribution of some molecules such as CN, C$_2$H, CO, or HCO$^+$ becomes narrower with increasing accretion rate, 
%GCG: while the distribution for other molecules remains nearly constant,
{ while other molecules present the opposite behavior (e.g., CS), and other molecules do not change their width under any circumstances (H$_2$CO, CH$_3$OH, NH$_3$)}.
{ Figure \ref{fig:fwhm} actually presents the molecules ordered with increasing filament width for the nominal accreting case. As can be seen from the figure, the molecules presenting smaller filament widths are also the molecules less sensitive to the different accretion rates. On the contrary, molecules with largest filament widths are also those most sensitive to accretion. This suggests that the formation of the group of `insensitive' molecules probably requires a particular density/temperature threshold and that the filament width simply indicates the width at which these physical conditions are reached, no matter the accretion history, and that the formation of the group of molecules sensitive to accretion is intimately related to the actual physical conditions of the gas.}

{ In order to quantify this a little bit more,}
in Table \ref{tab:sensitivity} we present the subset of the molecular species considered in this work, ordered by
sensitivity to the accretion rate.%
\footnote{Measured as
the absolute value of the logarithm of the ratio of filament widths measured for the nominal accretion to the static
models.}
{ The table shows that, as suggested already in Fig.~\ref{fig:fwhm}, our subset of molecules can be divided in two opposite groups, mainly the molecules `insensitive' to accretion, such as \ce{N2H+}, \ce{H2CO}, \ce{CH3OH}, \ce{NH3}, and the molecules `sensitive' to accretion such as CS, CN, \ce{C2H}, CO, and \ce{C3H2}. We classify a molecule as `insensitive' to accretion if the variation of its filament width with respect to the static case (in logarithmic scale) is smaller than $\sim10$\%, while those molecules with variations larger than $\sim40$\% will be considered as molecules `sensitive' to accretion. Note that some of the `sensitive' molecules have also been classified in previous works such as `early-time' molecules \citep[e.g.,][]{Taylor+98} or molecules tracing diffuse cores \citep[e.g.,][]{Frau+12}, while the `insensitive' molecules have been classified as `late-time' or tracing `deuterated cores'. However, the exact correspondence with previous works might not be perfect because, for example, here we do include grain-phase reactions which were not included by \citet[]{Taylor+98}.}

\begin{figure}
\centering
\includegraphics[width=0.49\textwidth]{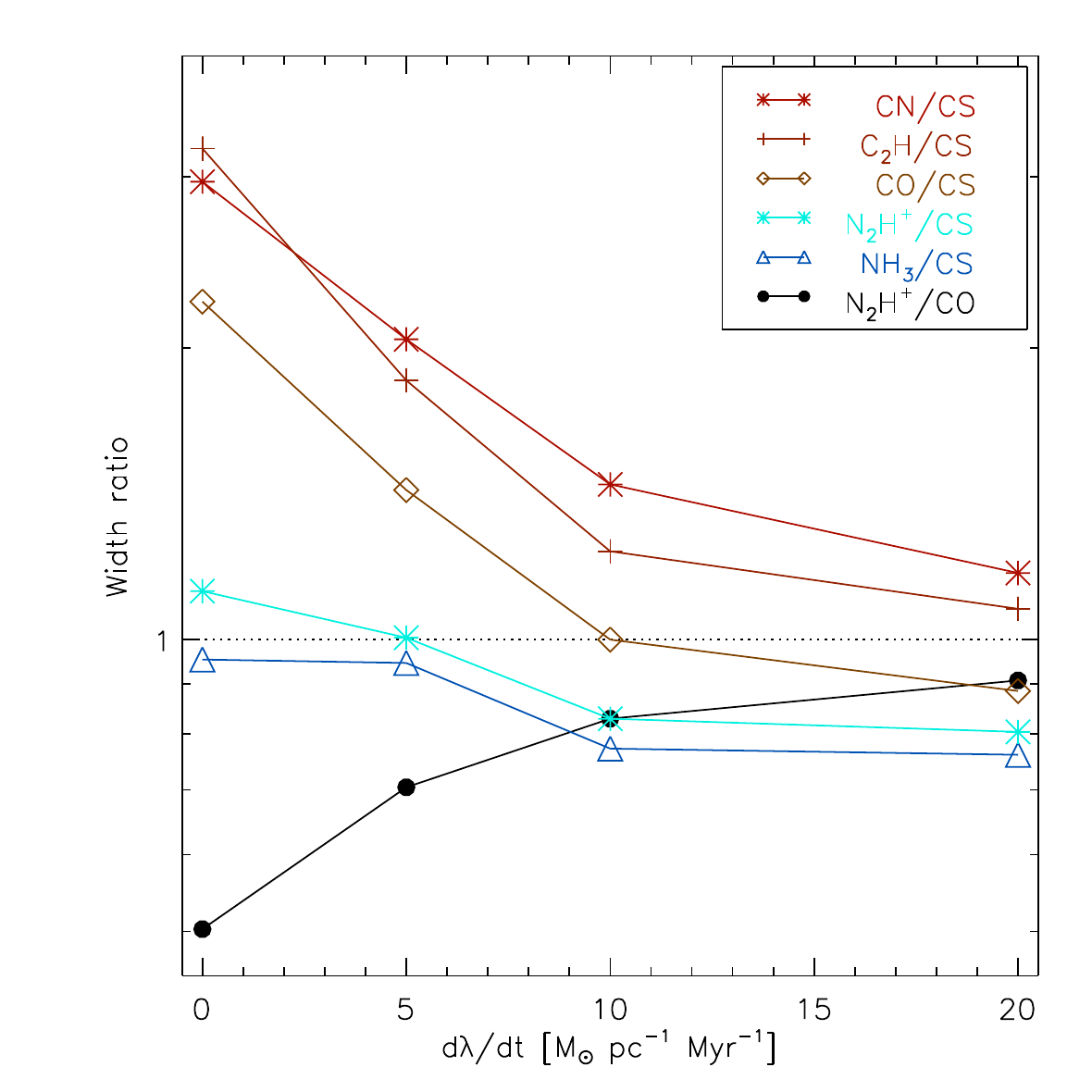}
\caption{Filament width ratios for CN, \ce{C2H}, CO, \ce{N2H^+}, and \ce{NH3}
to CS as a function of accretion rate to the filament.
%While the CN and C$_2$H distributions grow narrower with increasing accretion, the CS distribution becomes wider.
Colours and symbols for these molecular species correspond to those in Fig.~\ref{fig:abun}.
\ce{N2H+} to CO width ratio is also shown.}
\label{fig:ratios}
\end{figure}

The sensitivity to accretion rate may be (marginally) increased if we consider width ratios for pairs of molecular species.
Figure \ref{fig:ratios} shows the width ratios of CN and C$_2$H to CS as a function of accretion rate to the filament.
Although these ratios change by a factor of a few for the range of accretion considered, there is a clear trend
of decreasing width ratio of CN, C$_2$H, and CO to CS with increasing accretion.
{ As can be seen from Figure \ref{fig:ratios}, for large accretion rates the widths of the `insensitive' molecules tend to be the same and this might not be a strong test from an observational point of view. Thus, it might be also useful to study the width ratios between sensitive and insensitive molecules. For example, the width ratio of \ce{NH3} to CS, or the width ratio of \ce{N2H+} to CS should be smaller than one (CS should be always broader) for high accretion rates.}
%GCG: For other frequently observed species (N$_2$H$^+$ and NH$_3$, for example), this trend is not so clear but it is also present.

{ The results presented in Fig.~\ref{fig:ratios} indicate that observations of filaments with different molecular tracers could be a powerful tool to provide evidence of on-going accretion onto the filaments, if the appropriate tracers are used. In this sense, the observational work of \citet{Gong+18} can provide a first test to the model developed here. These authors observed multiple transitions in the Serpens filament using the Purple Mountain Observatory with a beam of about 50$''$, inferring a linear mass density of 36--41 M$_\odot$\,pc$^{-1}$, fully consistent with the linear mass density of the filaments of the GV14 simulations (of  $\sim40\,$M$_\odot\,$pc$^{-1}$). \citet{Gong+18} also measure the radial accretion rate onto the filament to be about 45 M$_\odot$\,pc$^{-1}$\,Myr$^{-1}$, only a factor of two larger than the accretion rate measured in the simulations of GV14, and similar to the accretion rate estimated by \citet{Palmeirim+13} and \citet{Shimajiri+19} in the Taurus main filament.
%the accretion rate of 45 is 72/1.6 taken from the abstract of Gong+18
Judging from Fig.~\ref{fig:ratios} and given the radial accretion rate measured in the Serpens filament, one would expect a width ratio of \ce{N2H+} to CS of about 0.8 and interestingly the fwhm of the CS distribution presented in Fig.2 of \citet{Gong+18} is 50\% larger than the \ce{N2H+} distribution, yielding a width ratio of \ce{N2H+} to CS of 0.7, very close to the value predicted in our model. 
However, opacity effects must be affecting the CS observations. \citet[]{Gong+18} also present abundance maps of C$^{18}$O and \ce{N2H+}. From Fig.~6 of \citet[]{Gong+18} we estimate a width ratio of \ce{N2H+} to C$^{18}$O of about 0.7. By comparing this value to the ratio shown in Fig.~\ref{fig:ratios} of \ce{N2H+} to CO (the behavior of C$^{18}$O should be very close to that of CO), we infer an approximate accretion rate of about 5~M$_\odot$\,pc$^{-1}$\,Myr$^{-1}$, assuming that the Serpens filament is similar to the filament taken from GV14.  
Although accurate works specifically designed to test this model should be carried out, the study of \citet{Gong+18} already constitutes a first test suggesting chemical differentiation due to accretion onto filaments.
%Furthermore, Fig.6 of Gong+18 provides abundance distributions of C18O and N2H+ showing a very striking difference between both molecules, with C18O being much wider. Since these two molecules are optically thin and sensitive to low temperatures, such a difference should be indicative of a difference in abundance. 
}

\subsection{Caveats of the model}

\ce{C2H} and CN are molecular species that are sensitive to the surrounding radiation field.
So, it could be argued that the
effect discussed here, if observationally verified, could be due to filaments embedded in different radiation fields.
In order to test this, we calculated the molecular abundances for the nominal accretion, for weaker (a factor of $0.3$)
and stronger ($\times 3$ and $\times 10$) radiation fields.
Although \ce{C2H} and CN abundances indeed change, the measured widths decrease only $80\%$, while other molecule
distributions become narrower by a similar amount.
More importantly, the CN to CS and \ce{C2H} to CS ratios remain relatively constant since almost all molecular species
also grow narrower with increasing radiation field.

For this study we used a specific filament density and velocity profile, namely the profile reported by GV14
for a simulation of a molecular cloud undergoing large-scale, hierarchical gravitational collapse.
GV14 fitted a Plummer profile with a core-radius of $0.11\pc$ for this filament's volume density distribution,
although the column density distribution is better fitted by a $0.31\pc$ core radius.
The corresponding fwhm for the column density distribution used here is $0.47 \pc$,
similar to the values obtained for every molecule, with exception of C$_2$H and CN,
which are the most sensitive to accretion rate.
This is not surprising since the molecular abundances relate in a complex fashion to the ambient density, even more
with a dynamically changing density.

%{\bf
Real filaments are not cylindrically symmetric.
Observations of filamentary structures in various molecular tracers frequently show elongated, velocity coherent, fiber-like substructures (\citealt{Hacar+13}, for example, but see also \citealt{Zamora+17}).
In the GHC scenario, these fibers may be seen as part of the large scale collapse, and so, the fibers will also be accreting from their environment, although at different density and accretion regimes to the ones explored in the model used here.
The effect described here, that different molecular tracers might yield different observed filament widths, should also happen at the fiber level.
Models exploring this in a variety of self-consistent flow conditions, including the lack of axial symmetry \citep{Naranjo20}, will be explored in a future contribution.
%}

As mentioned, the main characteristic of the flow around the filament is that the velocity smoothly turns from
radial to longitudinal, so that an accreted gas parcel slows down from supersonic to subsonic velocities
without the need of a shock.
In contrast, in the turbulent-compression model, filaments appear at the positions where flows meet and so we would
expect to find shocked gas around the filament, { with high abundances of typical shock tracers such as SiO or CH$_3$OH at the locus where the velocity turns from radial to longitudinal. Although widespread
SiO and CH$_3$OH emission has been found in some filamentary clouds \citep[e.g.,][]{JimenezSerra+10, DuarteCabral+14, Louvet+16, Cosentino+18, Li+20, Liu+20, Zhu+20}, the
turbulent-compression model requires to find them at high velocities, while these
shock tracers have been found associated with narrow linewidths of 1--6\,km\,s$^{-1}$.}
{ The presence of a shock at the filament border} should have an impact on the chemical abundances, { and detailed studies of shock tracers in the filament surroundings could be crucial} to distinguish these scenarios.

%{\bf
In the GHC scenario, the thermodynamic behavior of the gas around molecular filaments should have little effect on the gas dynamics since the gravitational collapse occurs in an almost pressureless fashion.
Still, \citet{Micic13} showed that the choice of cooling function might change the cloud morphology resulting in the simulation as slower cooling, shock-heated gas yields overpressured regions that expand against the thermal-instability induced flow.
Since the filament model under consideration corresponds to a later time, when gravitational instability drives the gas flow, the details of the cooling function should not have a large impact on our conclusions, but should be taken into consideration for models of filaments formed in the gravoturbulent scenario.
The simulations discussed by \citet{Micic13} show that using the cooling function from \citet[][as GV14 did]{KoyamaInutsuka02} yields dense regions that are $\sim 10\degK$ colder than their equivalent when the time-dependent cooling from \citet{Glover+07a,Glover+07b} is used.
With this in mind, we artificially increased the gas temperature by $10\degK$ in the filament model and repeated the calculation of filament width ratios for the static and nominal acretion cases.
We found that the ratios shown in Table \ref{tab:sensitivity} change by an average of $2\%$, which would be relevant only for molecular species that we classified as insensitive to accretion.
%}

%{\bf 
Finally, we note that we have post-processed the chemical evolution, i.e., we have decoupled the complex chemistry from the physical simulation. A fully self-consistent model would use the abundances from a coupled chemical model to compute the heating and cooling rates on the fly, and indeed this has been done but using reduced networks only \citep[see, e.g.,][for an example with simplified grain-surface chemistry]{Hocuk2016}.  
Such simplified networks focus on computing the abundances of key heating species and coolants at the expense of chemical complexity; however, we are interested in the abundances of key observables at sub-mm wavelengths which require much more complex networks in order to accurately (in as much is as possible) simulate their chemical evolution.  
However, it is worth to check if the results of the chemical simulations are consistent with the assumptions made in the heating and cooling functions used in GV14. The volatiles involved in the heating and cooling rates are \ce{H2} and CO, and atomic and ionic lines of CII, OI, FeII, and SiII \citep{Koyama2000}.  
Figure 1 in that work shows that at densities between $\sim 1$ ~cm$^{-3}$ and $\sim 10^{4}$~cm$^{-3}$ (relevant for this work) OI, CII, and CO are the primary coolants, and \ce{H2} formation/destruction is a primary heating source.  
\citet{Koyama2000} adopt elemental abundances of $4.6\times 10^{-4}$ and $3.0\times 10^{-4}$ for O and C, respectively. 
These values are similar to our assumed initial abundances shown in Table~\ref{tab:abundances}, albeit slightly higher.  
Hence the cooling rates would be expected to be slightly lower in a self-consistent version of our model although this slight change should not drastically change the gas temperature.  
We note that our \ce{H2} and CO abundance profiles with density differ from those presented in figure 1 of \citet{Koyama2000} in that we find molecular formation takes place at lower densities because we have a higher shielding column ($> 10^{21}$~cm$^{-2}$) than simulated in that work ($10^{19} - 10^{20}$~cm$^{-2}$).  
This is expected to increase the gas temperature at low densities due to increased heating provided by \ce{H2}; indeed, we find that is the case and at a density of $\sim 10$~cm$^{-3}$ \citet{Koyama2000} find a gas temperature of $\sim 100$~K whereas the gas temperature in the filament at this density is $\approx 300$~K. 
Hence, we conclude that the physical structure of the simulated filament should not be significantly affected by the decoupling of the hydrodynamics and chemistry; however, future simulations would be needed to fully quantify the differences.%}

The relation between widths measured for different molecules and accretion presented in Figs.~\ref{fig:fwhm} and
\ref{fig:ratios} is specific to the filament discussed in GV14.
While a fully self-consistent model of an accreting filament is beyond the scope of this contribution,
these results show a promising way to verify and measure accretion rates towards molecular filaments and it could be
applicable to other prestellar objects that may be modeled as out of equilibrium structures.

%%%%%%%%%%%%%%%%%%%%%%%%%%%%%%%%%%%%%%%%%%%%%%%%%%
\section*{Acknowledgements}

We thank E. Vázquez-Semadeni, J. Ballesteros-Paredes, and an anonymous referee for useful comments and suggestions.
%{\bf
G.C.G. acknowledges financial support from the UNAM-PAPIIT IN103822 grant.
%}
A.P. acknowledges financial support from the UNAM-PAPIIT IN111421 grant, and from the CONACyT project number 86372 of the `Ciencia de Frontera 2019’ program, entitled `Citlalc\'oatl: A multiscale study at the new frontier of the formation and early evolution of stars and planetary systems’, M\'exico.
A.P. and G.C.G. acknowledge support from CONACyT's Sistema Nacional de Investigadores.
%{\bf 
C.W.~acknowledges financial support from the University of Leeds, the Science and Technology Facilities Council, and UK Research and Innovation (grant numbers ST/T000287/1 and MR/T040726/1). %}

%%%%%%%%%%%%%%%%%%%%%%%%%%%%%%%%%%%%%%%%%%%%%%%%%%
\section*{Data availability}

The data underlying this article will be shared on reasonable request to the corresponding author.



\bibliographystyle{mnras}
\bibliography{2022_chemfilament} % if your bibtex file is called example.bib

%\begin{thebibliography}{99}
%\end{thebibliography}

%%%%%%%%%%%%%%%%%%%%%%%%%%%%%%%%%%%%%%%%%%%%%%%%%%

%\begin{figure*}
%\centering
%\includegraphics[width=\textwidth]{abun-conv.pdf}
%\label{fig:conv}
%\end{figure*}

% Don't change these lines
\bsp	% typesetting comment
\label{lastpage}
\end{document}